\documentclass[prl,twocolumn,superscriptaddress,showpacs,floatfix]{revtex4}
\usepackage{listings}
\usepackage{fancyvrb}
\usepackage{graphicx}
\usepackage{latexsym}
\usepackage{bm}
\usepackage{natbib}
\usepackage{dcolumn}
\usepackage{amsmath}
\usepackage{amstext}
\usepackage[english]{babel}
\usepackage{flafter}
\usepackage{url}
\usepackage{multirow}
\usepackage{float}
\newfloat{figtable}{t}{lop}
\floatname{figtable}{Table}

\newcommand{\st}{$^{\mathrm{st}}$}

\begin{document}

\title{Characterizing and modeling the dynamics of online popularity}

\author{Jacob Ratkiewicz}
\affiliation{School of Informatics and Computing, Indiana University, Bloomington, IN, USA}
\author{Santo Fortunato}
\affiliation{Complex Networks and Systems Lagrange Lab, Institute for Scientific Interchange, Torino, Italy}
\author{Alessandro Flammini}
\affiliation{School of Informatics and Computing, Indiana University,
  Bloomington, IN, USA}
\author{Filippo Menczer}
\affiliation{School of Informatics and Computing, Indiana University, Bloomington, IN, USA}
\affiliation{Complex Networks and Systems Lagrange Lab, Institute for Scientific Interchange, Torino, Italy}
\author{Alessandro Vespignani}
\affiliation{School of Informatics and Computing, Indiana University, Bloomington, IN, USA}
\affiliation{Complex Networks and Systems Lagrange Lab, Institute for Scientific Interchange, Torino, Italy}

\begin{abstract} 
Online popularity has enormous impact on opinions, culture, policy,
and profits. We provide a quantitative, large scale, temporal
analysis of the dynamics of online content popularity in two massive
model systems, the Wikipedia and an entire country's Web space.  We
find that the dynamics of popularity are characterized by bursts,
displaying characteristic features of critical systems such as
fat-tailed distributions of magnitude and inter-event time. 
We propose a minimal model combining the classic preferential
popularity increase mechanism with the occurrence of random  
popularity shifts due to exogenous factors. The model recovers the
critical features observed 
in the empirical analysis of the systems analyzed here, highlighting
the key factors 
needed in the description of popularity dynamics. 
\end{abstract}



\maketitle

The  dynamics of information and opinions have been deeply affected by the existence of 
Web-mediated brokers such as blogs, wikis, folksonomies, and search engines, 
through which anyone can easily publish and promote content online.
This ``second age of information''  is driven by the {\em economy of attention}, first 
theorized by Simon~\cite{Simon71Attention}.  Sources receiving a lot of attention become popular and have formidable power to 
impact opinions, culture, and policy, as well as advertising profit. 
The Web 2.0 and social media~\cite{wikinomics} not only modify
traditional communication processes 
with new types of phenomena, but also generate a huge amount of time-stamped data, making it possible for the first time to study the dynamics 
of online popularity at the global system scale. 

In this letter we focus on the dynamics of popularity of Wikipedia topics and Web pages. 
As popularity proxies we have chosen the traffic of a document, expressed by the number of clicks to that page generated 
by a specific population of users, and the number of hyperlinks pointing to a document. 
It is well documented that the statistical properties of these variables in the Web are very heterogeneous, 
with distributions characterized by fat tails roughly following power-law behavior~\cite{Albert99,Broder00,Meiss:2005jt,Meiss:2008cs}. Such distributions have
been explained with models based on the rich-get-richer mechanism~\cite{yule-simon, deSollaPrice76,ba+model}, 
but their validation from the point of view of the dynamical behavior
is problematic, mainly due to the difficulty to gather relevant data. 
The data sets utilized here, however, contain 
temporal information that makes it possible to observe the growth in popularity of individual 
topics or pages,  and allows us to statistically characterize the microdynamics by which online documents gather popularity.    

\begin{table}
\caption{Descriptions of the data sets constructed for our study. The two Wiki collections refer to indegree (1) and traffic (2) of Wikipedia topics, while the Chile collection refers to indegree of Chilean Web pages.}
\begin{center}
\begin{tabular}{l|ccccc}
\hline
 & & & Temporal \\
 & Vertices & Period & Resolution \\
\hline
{\bf Wiki$^1$} & 3,293,102 & Jan 2001 -- Mar 2007 & 1 sec. \\
{\bf Wiki$^2$} & 3,490,740 &  Feb 2008 -- Current & 1 hour \\
{\bf Chile} & 3,252,779 & 2001 -- 2006 & 1 year \\
\hline
\end{tabular}
\end{center}
\label{table:datasets}
\end{table}

Prior work on popularity dynamics has focused on news~\cite{Wu:2007xu,Dezso06}, 
videos~\cite{Szabo08, Crane:2008ud} and music~\cite{Salganik:2006et}.  
Here, we analyze three large scale data sets that we assembled 
about two information networks: the entire Wikipedia and the Chilean Web. 
Wikipedia is a large collaborative online encyclopedia with millions of articles 
and hundreds of thousands of registered contributors (\url{en.wikipedia.org}). 
By mining the full edit history of every article, we were able to reconstruct the entire 
Wikipedia structure at any past point in time. 
The raw data was available until March 2007 (\url{download.wikimedia.org}). 
Traffic data with hourly temporal resolution was obtained by cross-referencing 
with a separate data set originating from Wikipedia proxy server logs (\url{dammit.lt/wikistats}).
Our third data source is a yearly sequence of crawls of the Chilean Web, made available 
by courtesy of the TodoCL search engine (\url{www.todocl.com}). This data consists of 
one complete crawl of the \url{.cl} top-level domain for each of the years 2002--2006.  
Basic statistics on each data set are shown in Table~\ref{table:datasets}. 
The representative graphs of these data sets have  
an approximately power-law distribution of indegree~\cite{chilean+web,
capocci+pref+attach+wiki, zlatic+wiki}, like the Web graph at large.

In order to gauge quantitatively the popularity of documents  we consider the number of hyperlinks pointing to a page (indegree $k$ in the graph representation of the Web~\cite{Albert99}), and the traffic $s$ of the page, expressed by the number of clicks to it. Given either of these two popularity proxies $x_{t}$ at time $t$, we study its 
\emph{logarithmic derivative} $[\Delta x/x]_t = (x_t - x_{t-1}) / x_{t-1}$,
which represents the relative variation of the measure in the time unit.

\begin{figure}
\centerline{\includegraphics[width=\columnwidth]{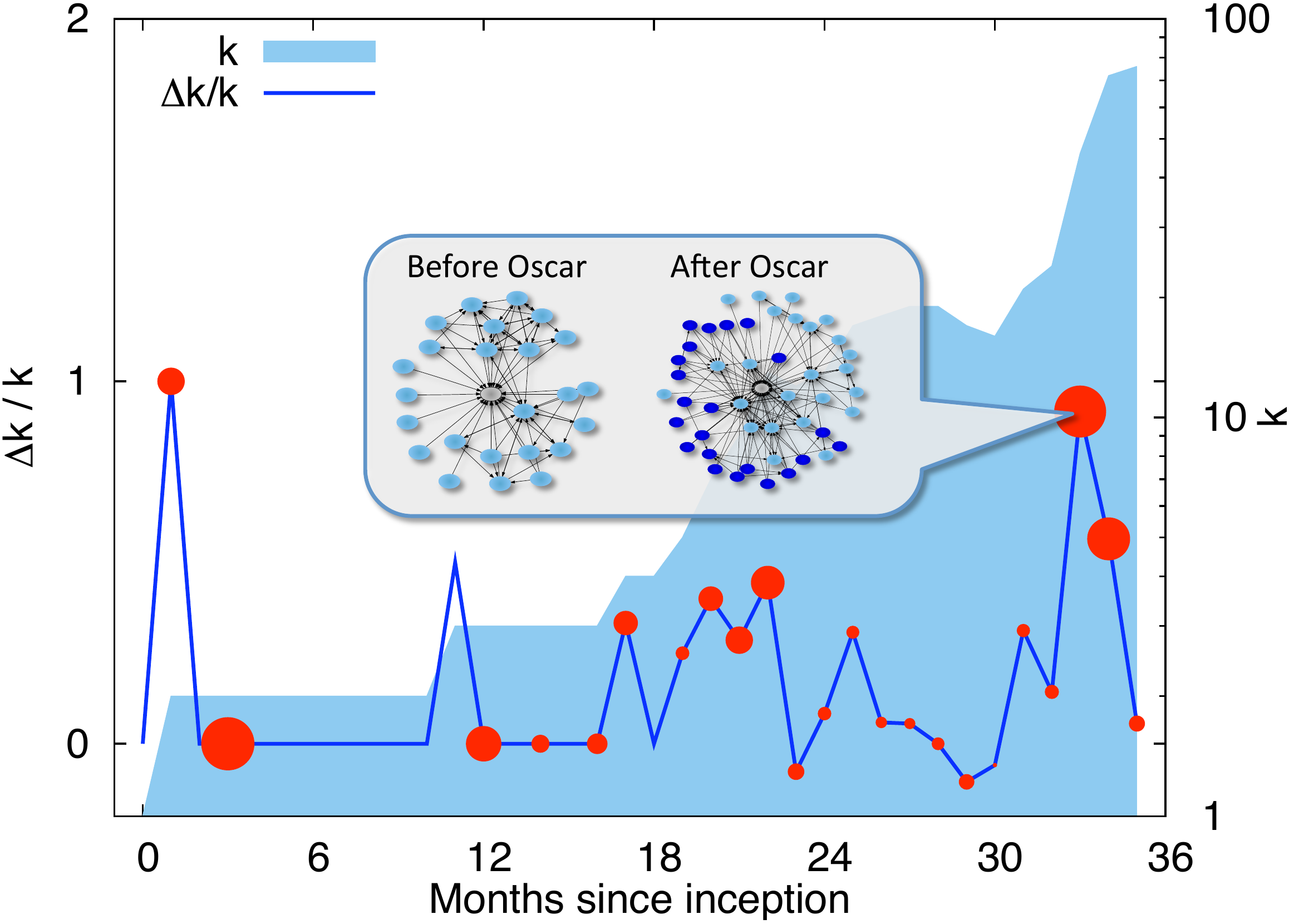}}
\vspace{-1em}
\caption{\label{fig:bursty-pages} 
Time series of indegree $k$ and its logarithmic derivative $\Delta
k/k$ for the Wikipedia topic page about the artist Jennifer Hudson.
Topics typically experience a burst in their early life.  Here we observe later fluctuations as well. 
Jennifer Hudson became popular through a television show leading to her first burst. 
Another occurred when she won an Academy Award; degree popularity doubled as many other pages 
linked to the article (inset). The size of each circle shows another popularity measure; it is proportional to the log-derivative of the number of times the article 
is revised. The article receives more edits when it attracts more links.
}
\end{figure}

Fig.~\ref{fig:bursty-pages} shows the logarithmic derivative of the
indegree \emph{vs} time for an 
example page in the English Wikipedia. Despite a roughly exponential
growth, the logarithmic 
derivative provides a signature by which different topics can be
compared on the same scale. 
Almost all pages experience a burst in $\Delta x / x$ near the
beginning of their life. 
Many pages receive little attention thereafter. While some pages
maintain a nearly 
constant positive logarithmic derivative indicating an exponential
growth, a number of pages continue to experience intermittent bursts in $\Delta x/x$ later in their life as in the example.

\begin{figure}
\includegraphics[width=4.2cm]{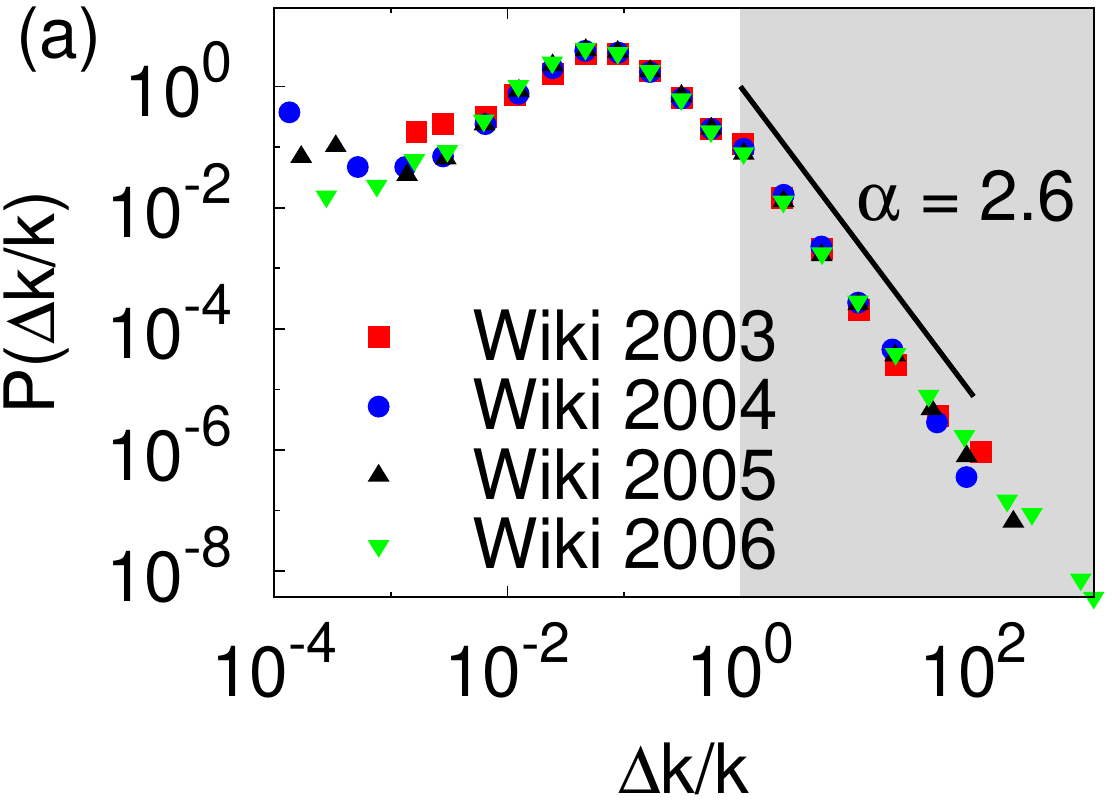}
\includegraphics[width=4.2cm]{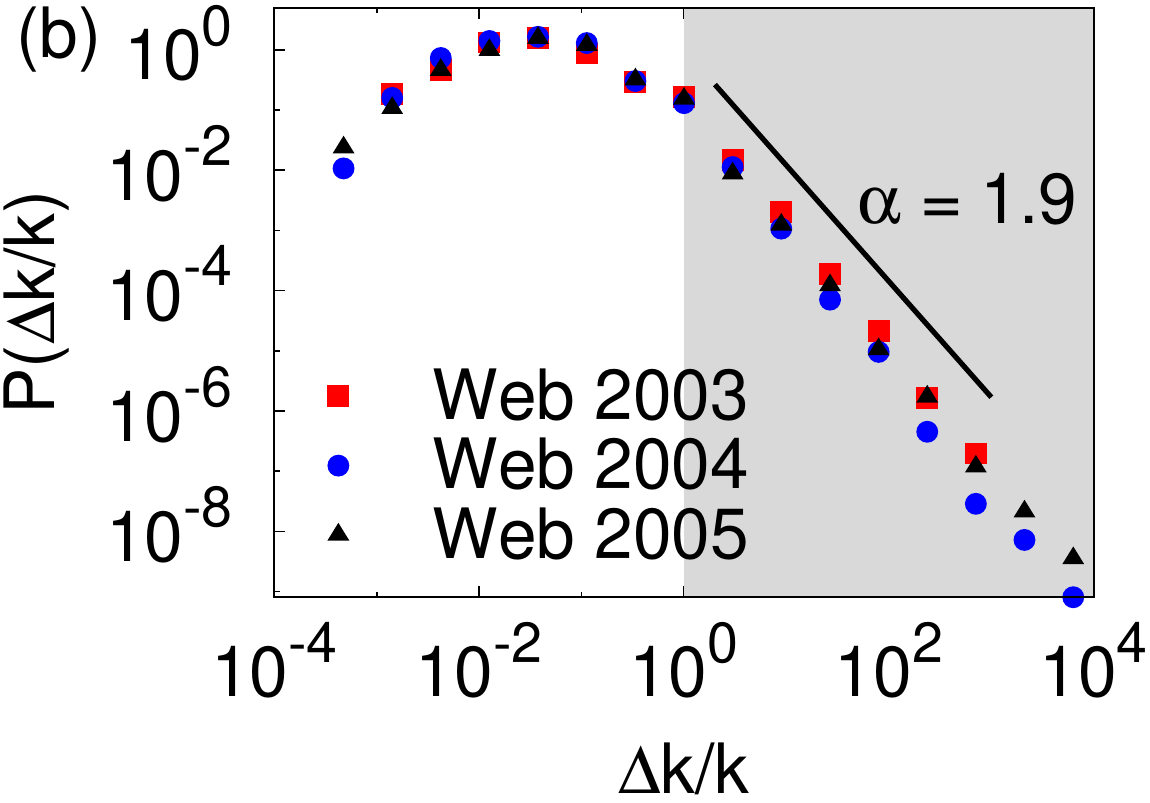}
\includegraphics[width=4.2cm]{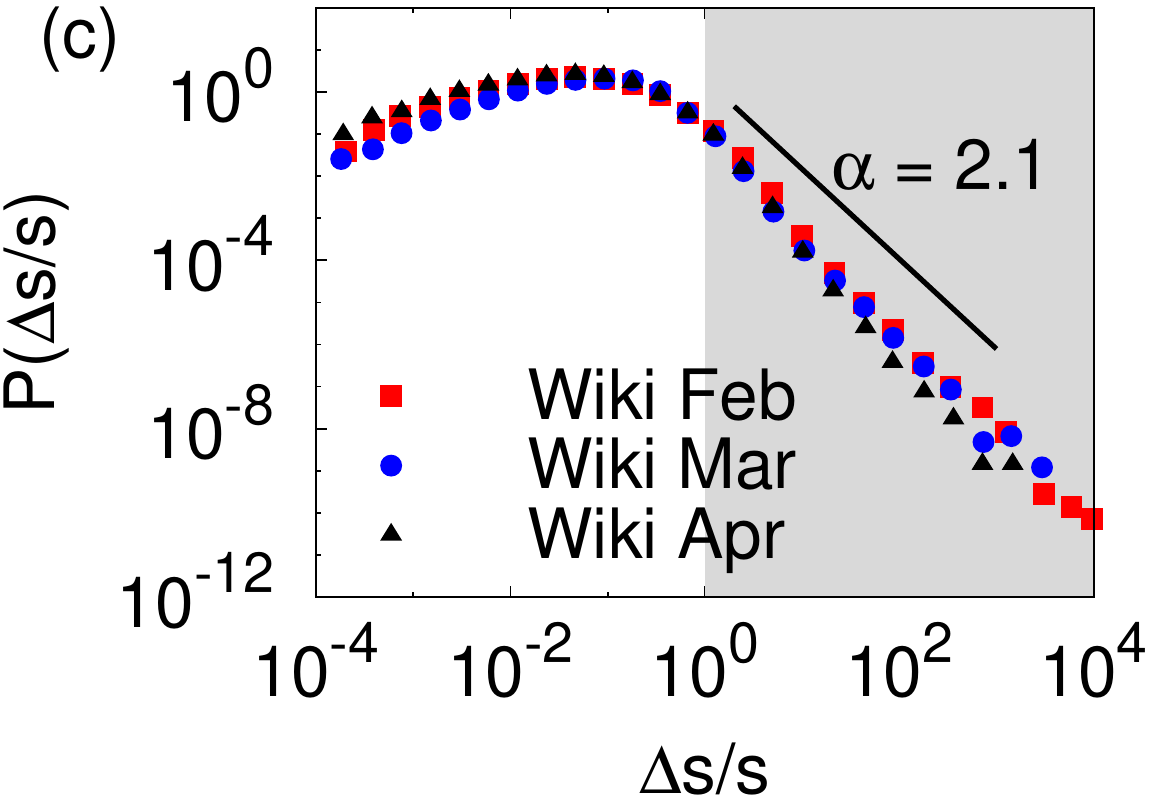}
\includegraphics[width=4.2cm]{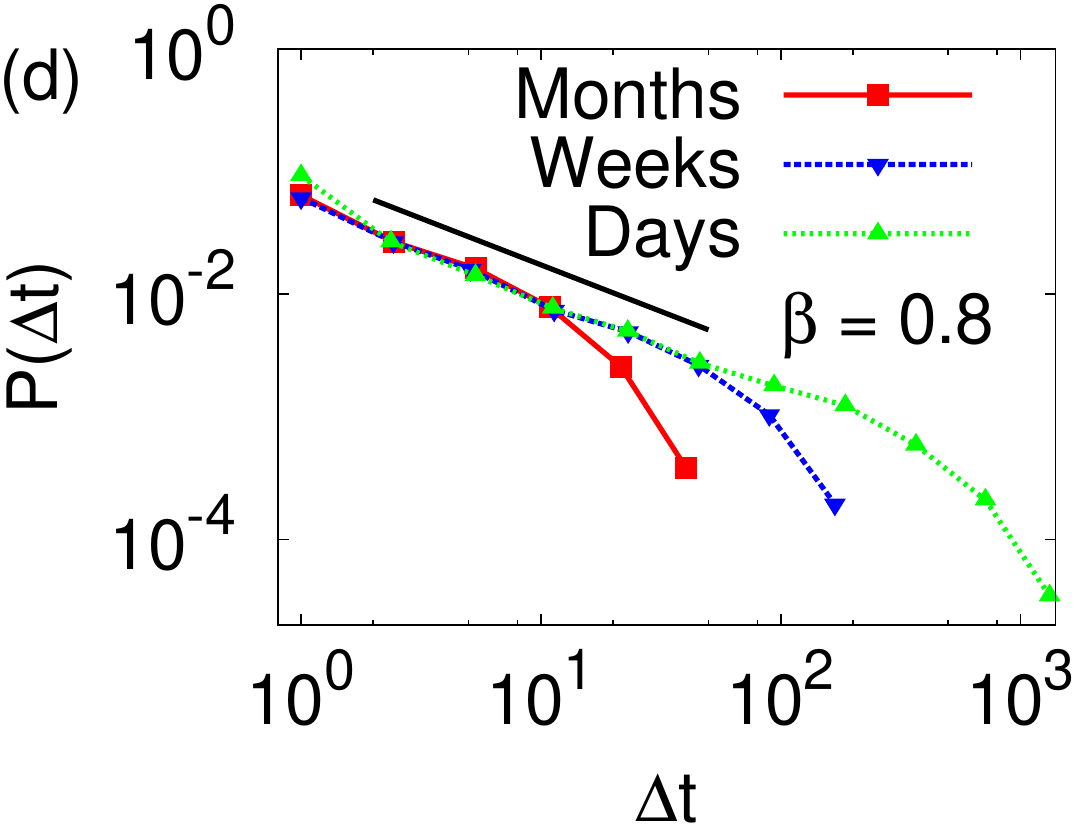}
\vspace{-1em} 
\caption{\label{fig:delta}
(a, b, c) Distributions of popularity burst size. The gray areas highlight the 
events for which $\Delta k>k$ (hence $\Delta k/k>1$). Maximum likelihood methods~\cite{Clauset07powerlaws} in conjunction with the 
Kolmogorov-Smirnoff (KS) statistic rule out lognormal fits. 
In each case the KS statistic suggests that the
power-law curve is the better fit for the tail. For the distribution of $\Delta k/k$ in Wikipedia (a)  
the parameters are $\alpha = 2.6$ for the exponent of the power law, with a lower cutoff of $12$ and a KS statistic of $0.005$.
For the Web (b) we find $\alpha = 1.9$ for the exponent of the power law, with a lower cutoff of $42$ and a KS statistic of $0.007$.
For the distribution of $\Delta s/s$ the parameters are $\alpha = 2.1$
with lower cutoff $90$ and KS statistic  $0.007$. The slopes of the
best fit power laws are shown as guide to the eye.
These behaviors are consistent 
across a wide range of temporal resolutions, as observed using time units from a day to a year.
(d) Distribution of the time interval $\Delta t$ between consecutive indegree bursts of Wikipedia articles. 
We consider bursts such that $\Delta k/k > 1$ after January 1\st, 2003. 
The three curves correspond to different time resolutions of months, weeks, and days, aligned on the 
$x$-axis for ease of visualization. As we increase the resolution the tail of the distribution extends 
further, an indication that the cutoff is a finite size effect. As a guide to the eye we show a power 
law $P(\Delta t) \sim (\Delta t)^{-\beta}$ with $\beta \approx 0.8$.
}
\end{figure}

The distribution of magnitude $\Delta x / x$ for the two popularity measures at representative 
time resolutions is illustrated in Figs.~\ref{fig:delta}a--c. In all cases and at 
all granularity we observe a heavy-tail behavior.
Such heavy-tailed burst magnitude distributions suggest a dynamics lacking a characteristic 
scale. This is typical in a wide range of ``critical'' physical, economic, and social systems, 
such as avalanches, earthquakes, stock market crashes 
and human communication~\cite{Barabasi05bursty, Mandelbrot97Fractals, 
Stanley:1996it, gut:44,rybski09}.  Further evidence comes from the study of the 
distribution of the length of inter-event intervals. 
For each document we record the time stamp of each event 
for which $\Delta x/x>1$ and measure the inter-event times $\Delta t$. 
The probability distributions of $\Delta t$ in the different data sets 
(Fig.~\ref{fig:delta}d) are not distributed following a
Poissonian, as expected by queueing theory in traditional systems, but 
in a power-law fashion with a finite size cutoff, 
as in Omori's law of earthquakes~\cite{omori:894} and other
self-organized criticality phenomena~\cite{bak87}.

\begin{figure}
\includegraphics[width=4.2cm]{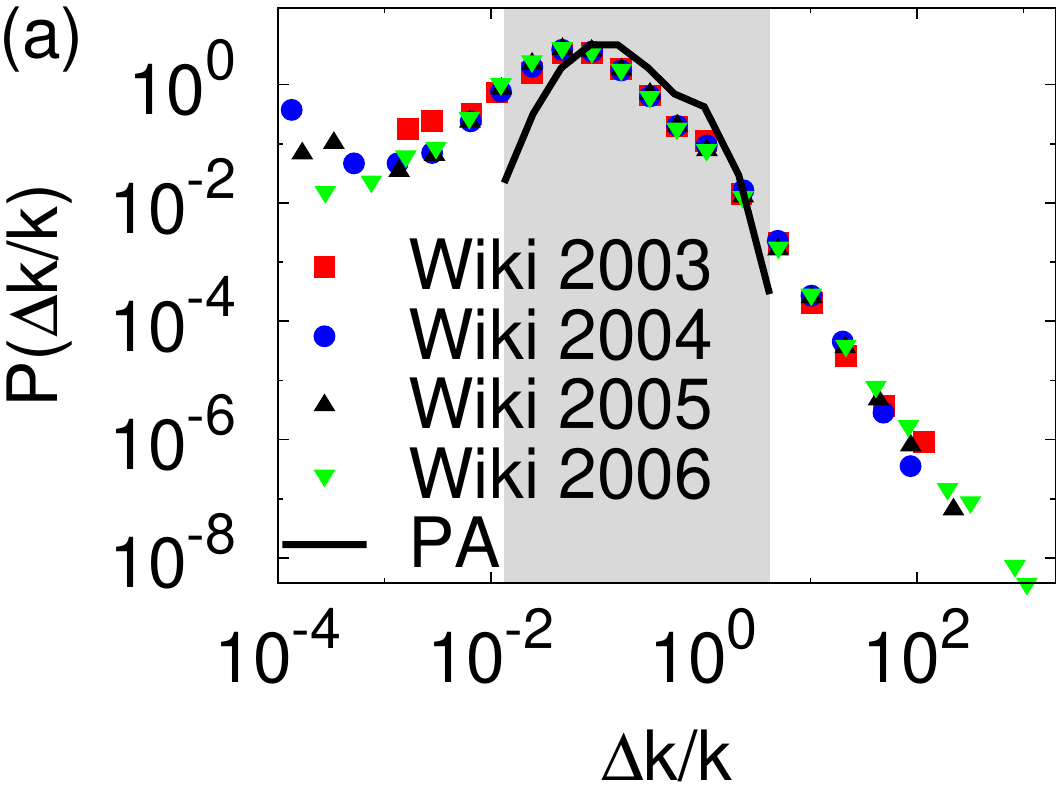}
\includegraphics[width=4.2cm]{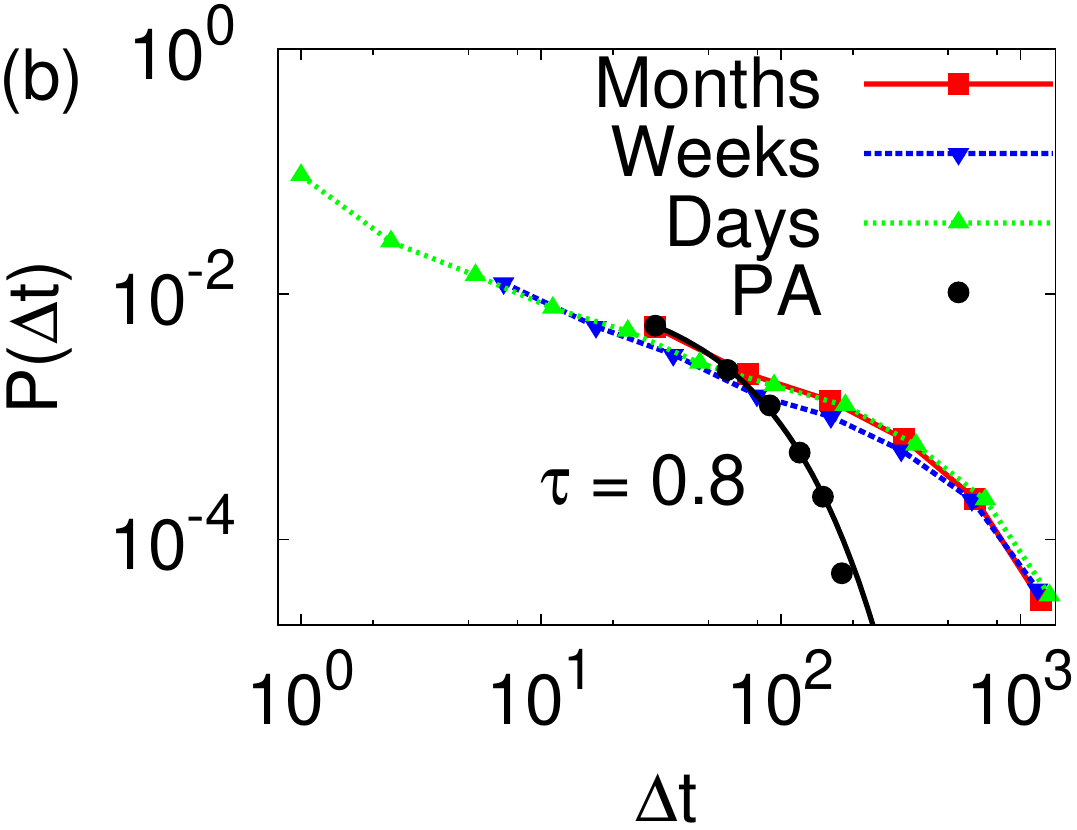}
\includegraphics[width=4.2cm]{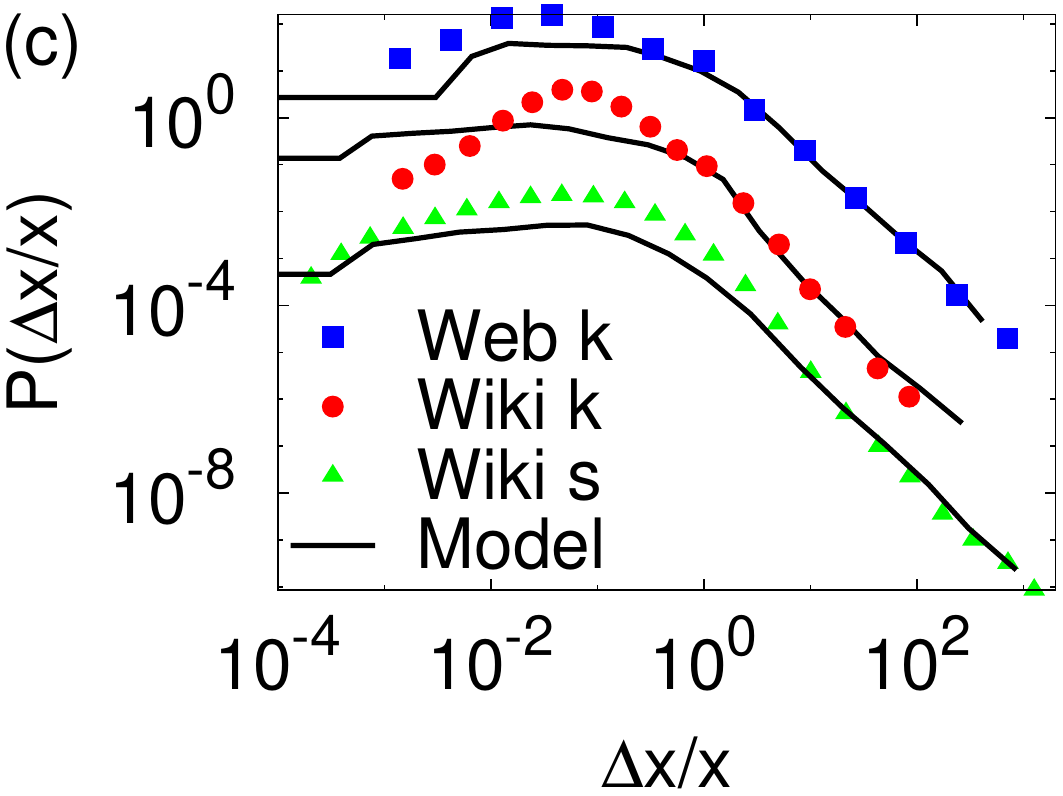}
\includegraphics[width=4.2cm]{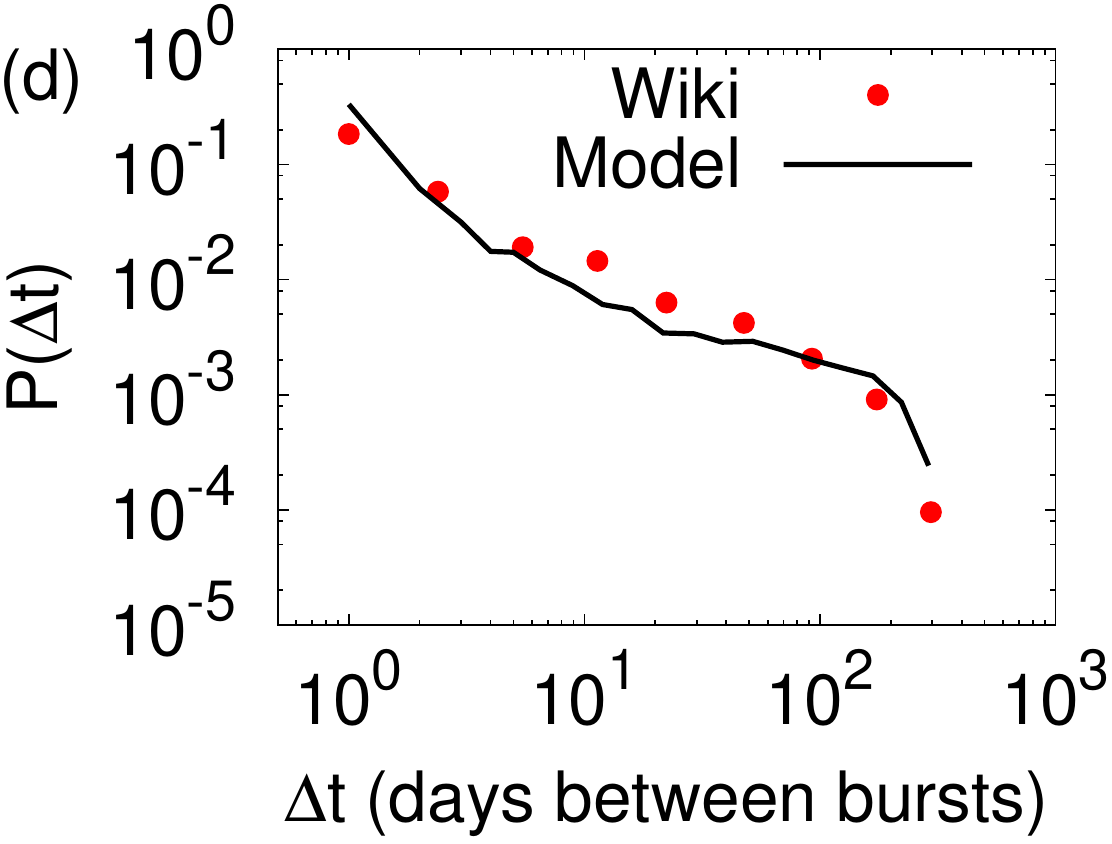}
\vspace{-1em}
\caption{\label{fig:model}
(a) Comparison of the empirical burst size distributions with what
would be expected from a preferential attachment (PA)
process. Extensive numerical tests and maximum likelihood
fitting~\cite{Clauset07powerlaws} show that PA generates an approximately lognormal distribution (defined inside the gray area) inconsistent with the long tail observed in the empirical data. 
(b) The empirical inter-burst time distributions overlap when
time is expressed in terms of the same unit (in the figure, the common
time unit is one day). The distribution generated by PA is much narrower and fits an exponential $P(\Delta t) \sim e^{-\Delta t/\tau}$ with $\tau = 0.8$.
(c,d) The rank-shift model, despite its simplicity, reproduces quite well the distributions of both 
event size (c) and inter-event time (d).}
\end{figure}

The clear evidence for the bursty behavior of online popularity dynamics calls for a 
stylized model able to explain the observed features in terms of the
already acquired popularity of each page and the shifts in collective
attention triggered by exogenous events. 

\begin{figure}
\includegraphics[width=4.2cm]{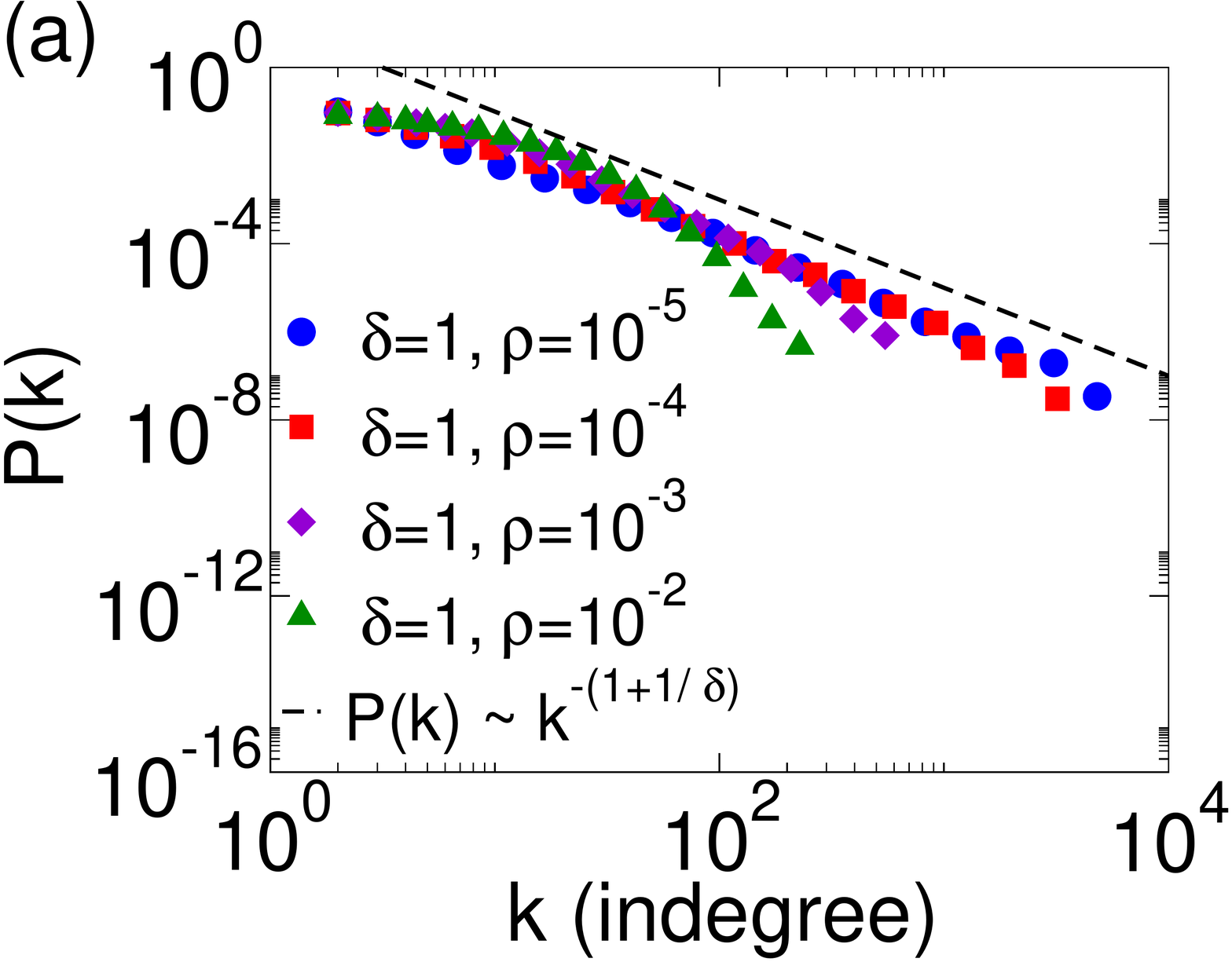}
\includegraphics[width=4.2cm]{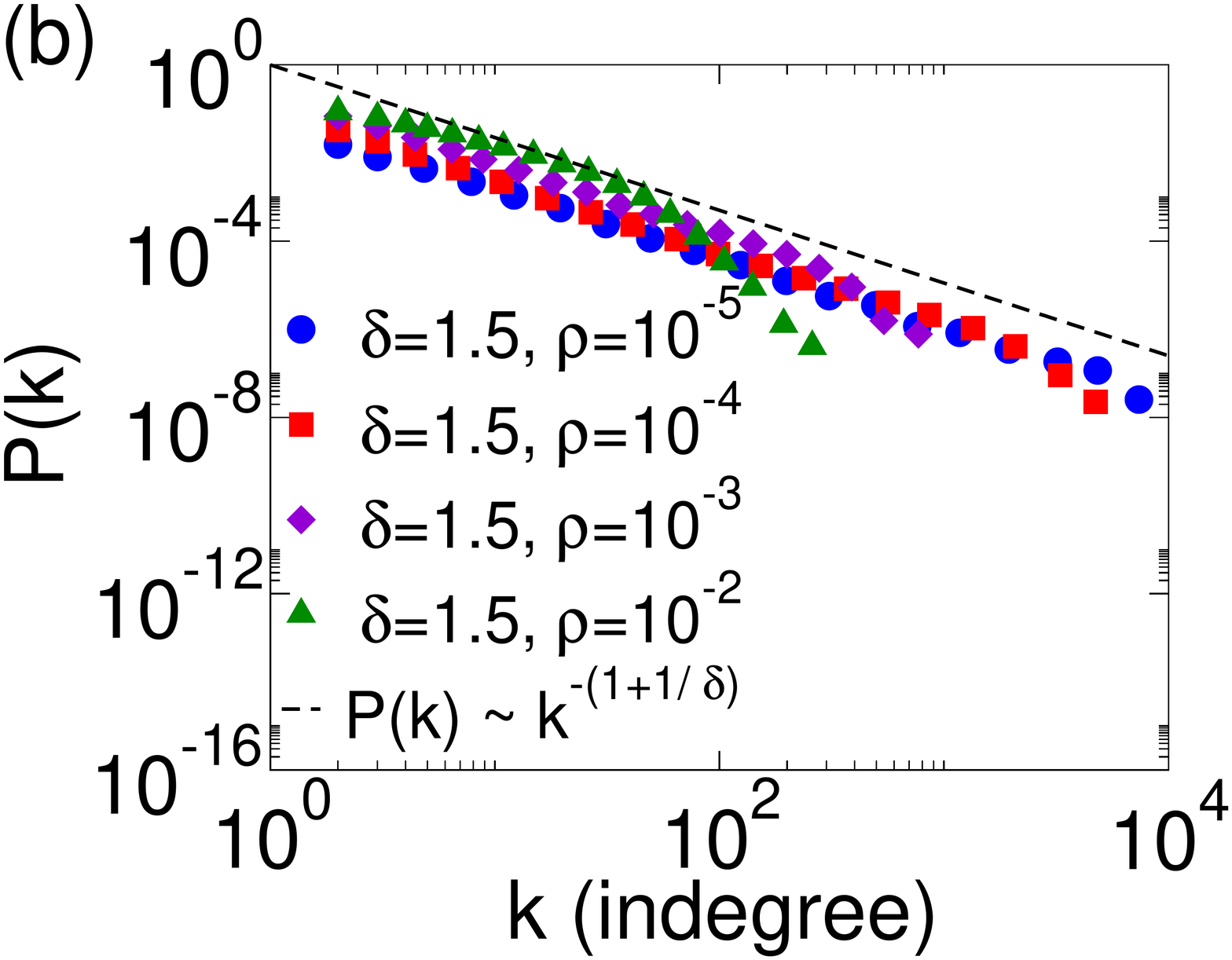}
\includegraphics[width=4.2cm]{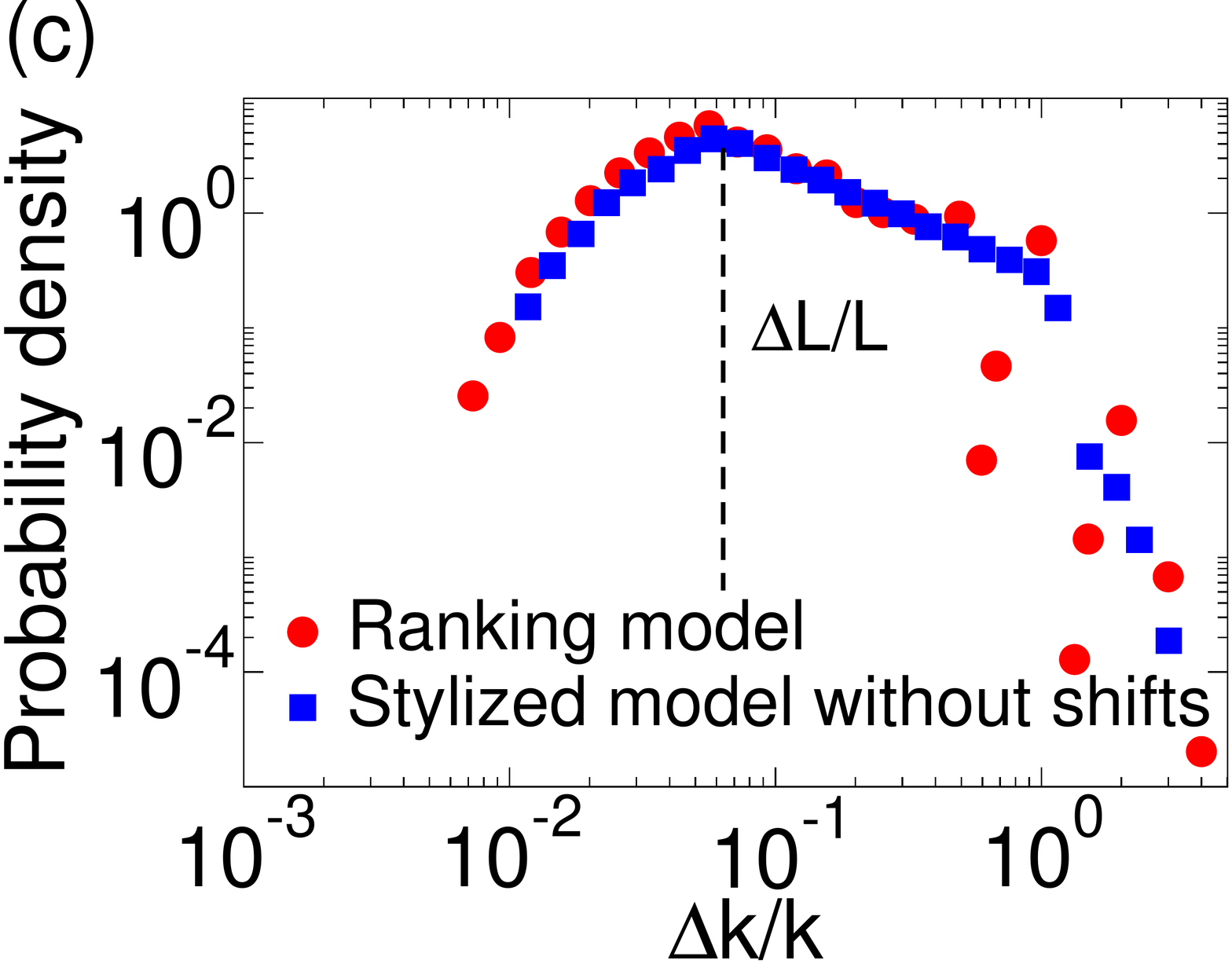}
\includegraphics[width=4.2cm]{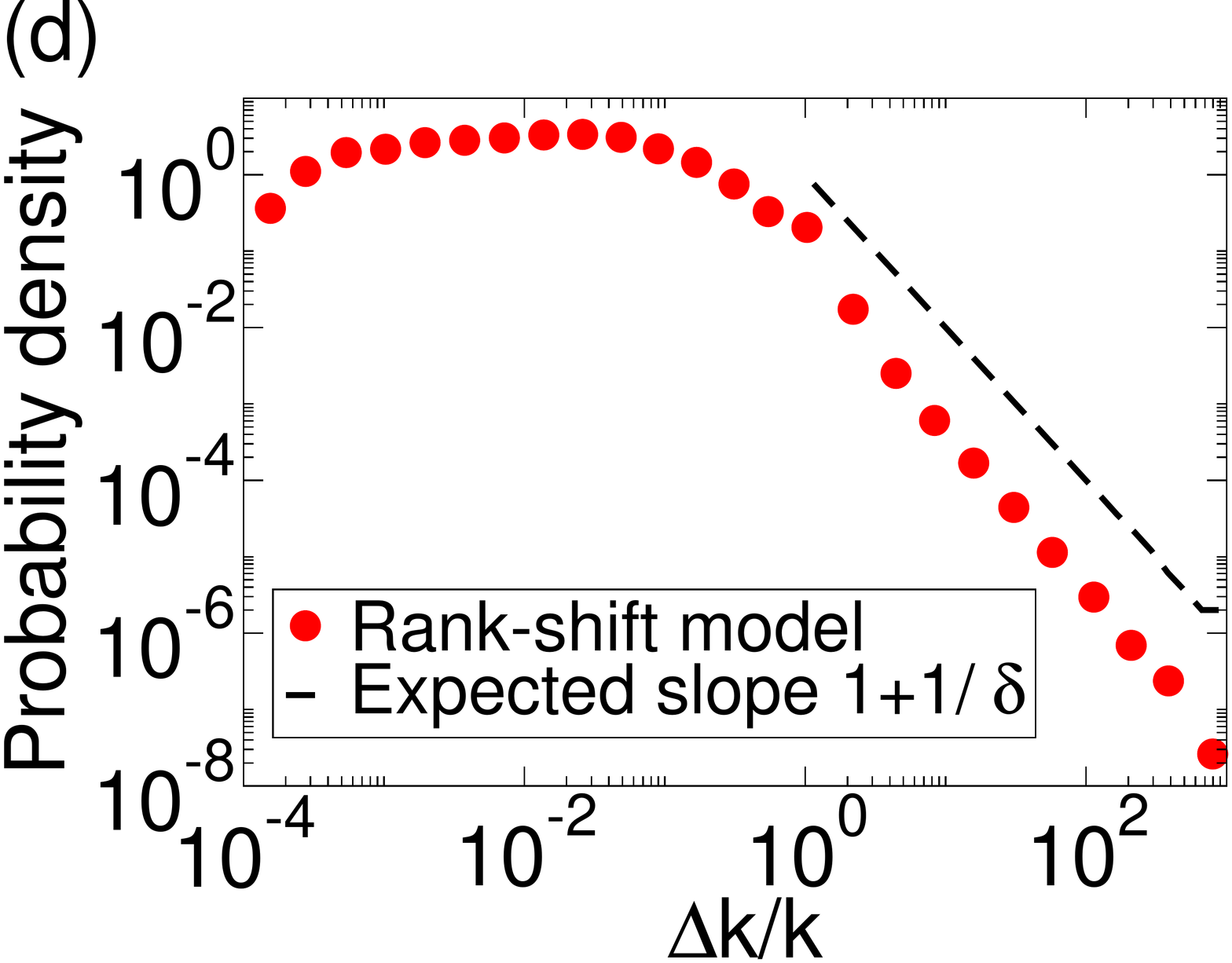}
\vspace{-1em} 
\caption{\label{fig:model0}
Rank-shift model. (a), (b). Indegree distribution: $\delta=1$ (a),
$\delta=1.5$ (b). (c) Comparison of the distribution of popularity
bursts for the ranking model~\cite{fortunato+ranking+growth}
(circles) and a stylized model built upon the simple assumptions of
growth described in the text. (d) Comparison of the distribution of
popularity bursts with the expected slope derived by assuming that nodes are reranked at most once.}
\end{figure}

The rich-get-richer mechanism can be simulated with the classic linear
preferential attachment model~\cite{ba+model}, in its directed version~\cite{Doro00}, 
or with the \emph{ranking model} by  Fortunato 
\textit{et al.}~\cite{fortunato+ranking+growth}. 
In the latter items are ranked according to their
popularity $x$, and the probability that an existing item $i$ receives
a unit (e.g., a click)  is $P(i) \sim r_i^{-\delta}$, where $r_i$ is
the rank of $i$ and $\delta>0$ is a free parameter that tunes the
power-law popularity distribution $P(x) \sim x^{-\gamma}$, such that
$\gamma = 1 + 1/\delta$. Both preferential attachment and ranking
models, however, fail to reproduce the long tails observed in the
distributions of both $\Delta x/ x$ and $\Delta t$ (Figs. \ref{fig:model}a-b). 
Neither model accounts for the occurrence of exogenous factors that
shift the attention of users and suddenly 
increase the popularity of specific topics because of events such as an
actor winning a prize, political elections, etc. 
The minimal assumption in modeling exogenous perturbation consists in considering external stochastic events 
interfering
with the basic rich-get-richer mechanism by 
suddenly changing the popularity of a topic. The simplest way to implement this mechanisms consists in 
introducing in the ranking model  a {\it reranking
probability} $\rho$, 
such that at each iteration every item is moved to a new position toward the 
front of the list, chosen randomly with equal probability between 1 (the top 
position) and the node's current rank $j$. We call this the {\it rank-shift model}~\cite{epaps}.  

In Fig.~\ref{fig:model0}a and \ref{fig:model0}b we show the indegree
distribution of the rank-shift model
for several values of $\rho$: $\delta=1$ (a) and
$\delta=1.5$ (b). The ranking model ($\rho=0$) yields the slope
$1+1/\delta$ indicated by the dashed line. The reranking
probability introduces an exponential cutoff in the distribution,
which becomes relevant for $\rho \approx 10^{-2}$ and larger (but we used 
$10^{-5} < \rho < 10^{-3}$ in our simulations).

The distribution of $\Delta k/k$ 
shows two distinctive features, which are remarkably found in the
empirical distributions: a maximum located in the range 0.01--0.1
and a fat tail. Since the reranking probability is low, to understand the existence and the location of the
maximum it is convenient to consider the model in the absence of the
reranking mechanism. 
At a large time $T$, the expected value of the
degree of the node with rank $r$ is proportional to $Lr^{-\delta}$,
where $L$ is the number of links present in the network at time $T$.
Let $\Delta L$ be the number of links added during the interval
$\Delta T$ at whose extremes the ratio $\Delta k/k$ is computed. Let
$\Delta L\ll L$, an assumption verified in our
calculations. Therefore, one can safely assume that in the period
$\Delta T$ the addition of new links does not affect significantly the
degree of nodes and their relative ranking. So one can regard the growth process as a multinomial
process with probabilities $p(r)\propto r^{-\delta}$. The expected
number $\Delta k$ of new links acquired by a node of rank $r$ is
therefore $p(r)\Delta L$. 
The assumption of (almost) stationarity also
provides that $k(r)\sim p(r)L$. We therefore expect $\Delta k/k$
for a node to be distributed around $\Delta L/L$, regardless of the node.
In Fig.~\ref{fig:model0}c we compare the simulation of
the ranking model with the one of the multinomial process with
$p(r)\propto r^{-\delta}$, by using the parameters relative to the
Wikipedia data set of January 2003, which represents an ideal tradeoff
between the needs of having a sufficient number of bursts and a system
size not too large for the model to run.
The number of nodes/pages was
$N\approx 1.3\cdot 10^5$, the number of hyperlinks $L\approx 1.3\cdot
10^6$ and $\Delta L \approx 8 \cdot 10^4$. Based on the above discussion
we expect to observe a maximum in the distribution of $\Delta k/k$
located at $\Delta L/L\approx 0.06$. This is exactly where
the maxima of the empirical distributions of popularity bursts are
located (see Fig. 2a).

The ranking model cannot reproduce
the fat tail observed in the real data. This is the reason why we
introduced the reranking mechanism in our model. Here, it is the
nodes that are suddenly promoted to a higher rank that are responsible
for the high values of $\Delta k/k$ in the simulations. 
We consider a node that at time $T$ (the reference time
at which we start measuring $\Delta k$) has rank $r_1$, and is
immediately promoted to rank $r_2$, with $r_2$ chosen uniformly in
$1\leq r_2\leq r_1$. Under the same assumption of stationarity that we
made above, the expected degree of the node before promotion is
$k(r_1)\approx Lp(r_1)\propto r_1^{-\delta}$. Let us further assume
that $\rho \ll 1$ and that $\Delta L \ll L$, 
which hold for the parameters used in our model. 
Since the reranking probability is small, we can safely
assume that no node is reranked more than once during the observation
time $\Delta T$. The expected number of links collected during the
period $\Delta T$ is then $\Delta k=\Delta Lp(r_2)\propto
r_2^{-\delta}$. We expect therefore $\Delta k/k\propto (r_2/r_1)^{-\delta}$.
It is straightforward 
to derive the distribution $P(\Delta k/k)$ for a
generic node that is promoted at the beginning of $\Delta T$ by
considering all pairs of values $r_1$, $r_2$ uniformly distributed in
$1\leq r_2\leq r_1\leq N$. We find $P(\Delta k/k) \propto (\Delta
k/k)^{-(1+1/\delta)}$. In Fig.~\ref{fig:model0}d we highlight the tail of the
distribution $P(\Delta k/k)$ as produced by the rank-shift model and
our expectation for its slope: the match is surprisingly
good.

Simulations of the rank-shift model were performed using 
parameters matching those from the empirical data (e.g., $N=2.8 \times 10^{5}$ nodes for the 
Wikipedia in 2003); the free model parameters were set to fit the empirical distributions: 
$1 \leq \delta \leq 1.2$ and $10^{-5} \leq \rho \leq 10^{-3}$. 
For $\rho=0$ we recover the original ranking model, which yields a
lognormal distribution of $\Delta x/x$, 
like the preferential attachment (Fig. \ref{fig:model}a). For $\rho>0$ numerical simulations show that the tail 
of the popularity burst magnitude distribution shifts from a lognormal to a power law. The popularity 
distribution itself remains a power law; its exponent remains $\gamma=1+1/\delta$, but with an exponential cutoff  
depending on $\rho$.

Such a parsimonious model is able to reproduce the most relevant features observed in the empirical data. 
Not only does rank-shift predict the distributions of both popularity measures in our data sets, 
but also the long tails of the distributions of 
indegree and traffic burst size (Fig.~\ref{fig:model}c). 
Furthermore, it naturally accounts for the maxima of the empirical distributions. 
Remarkably the model captures the long-range distribution of inter-burst intervals as well (Fig.~\ref{fig:model}d). 
The random rank-shift mechanism is therefore able to capture the way in which Web sites and pages gain and 
accumulate popularity: not by a gradual proportional process, but by a sequence of bursts that move 
them to the forefront of people's attention. 
Such bursts  are  
different from those observed in news-driven events~\cite{Wu:2007xu}, where attention fades 
rapidly and overall popularity is lognormal-distributed. 
We also found that smaller rank shifts are unable to capture 
the critical burst behavior observed in the data~\cite{epaps}.

At the present stage our model is mostly descriptive and simply aims at reproducing at the coarsest 
level the distributions that characterize popularity changes. Possible
refinements may include the effect of
search engines, external events, news, word of mouth, social media, marketing campaigns, 
or any combination of them. The study of traffic patterns and models~\cite{Meiss:2008cs, Goncalves09wsdm, Meiss10HT} 
may help shed empirical light on this question. 


\begin{acknowledgments}
We thank R.~Baeza-Yates, C. Cattuto, B.~Dravid, V.~Griffith,
V. Loreto, M. Marchiori, M.~Meiss.
This work was supported in part by a Lagrange Senior Fellowship from the CRT Foundation to F.M., NSF grant IIS-0513650 to A.V., and the Lilly Endowment Foundation. S.F. gratefully acknowledges ICTeCollective, grant 238597 of the European Commission.
\end{acknowledgments}

\end{document}